\documentclass[a4paper]{jpconf}
\usepackage{graphicx}
\usepackage{color}

\begin{document}
\title{Noncoplanar spin canting in lightly-doped ferromagnetic Kondo lattice model 
on a triangular lattice}

\author{Yutaka Akagi and Yukitoshi Motome}

\address{Department of Applied Physics, University of Tokyo, Hongo 7-3-1, Bunkyo-ku, Tokyo 113-8656}

\ead{akagi@aion.t.u-tokyo.ac.jp}

\begin{abstract}
Effect of the coupling to mobile carriers 
on the 120$^\circ$ antiferromagnetic state is investigated 
in a ferromagnetic Kondo lattice model on a frustrated triangular lattice. 
Using a variational calculation for various spin orderings up to a four-site unit cell, 
we identify the ground-state phase diagram with focusing on the lightly-doped region. 
We find that an electron doping from the band bottom immediately destabilizes 
a 120$^\circ$ coplanar antiferromagnetic order and induces 
a noncoplanar three-sublattice ordering  
accompanied by an intervening phase separation. 
This noncoplanar phase has an umbrella-type spin configuration 
with a net magnetic moment and a finite spin scalar chirality. 
This spin-canting state emerges 
in competition between the antiferromagnetic superexchange interaction and 
the ferromagnetic double-exchange interaction under geometrical frustration.  
In contrast, a hole doping from the band top retains the 120$^\circ$-ordered state 
up to a finite doping concentration and does not lead to a noncolpanar ordering. 
\end{abstract}

\section{Introduction}
Coupling between charge and spin degrees of freedom is a fertile ground 
for various fascinating phenomena in strongly correlated electron systems. 
One of the fundamental models for describing the interplay 
is the ferromagnetic Kondo lattice model, or 
equivalently, the double-exchange (DE) model~\cite{Zener1951}. 
The model includes the ferromagnetic Hund's-rule coupling 
between itinerant electrons and localized spins as well as
the antiferromagnetic (AF) superexchange (SE) interaction between localized spins. 
In addition to the early researches~\cite{Zener1951,Anderson1955,de-Gennes_1960}, 
the rediscovery of the colossal magnetoresistance (CMR) effect 
has stimulated intensive studies and many interesting properties have been revealed~\cite{Kaplan1999,Tokura1999,Dagotto_2001}. 
Particularly, in the lightly-doped regions, 
intricate behaviors show up because of the keen 
competition between 
the AF SE interaction and 
the ferromagnetic DE interaction induced by mobile carriers. 
In the early stage of the study, de Gennes predicted a spin canting state with 
a spin-flop type ordering, which smoothly connects 
collinear AF ordering in the undoped insulator and ferromagnetism in a doped metal~\cite{de-Gennes_1960}. 
Later, the scenario has been revisited;  
it was argued that a phase separation (PS) takes place to hinder the canting state and 
play an important role in the CMR effect~\cite{Dagotto_2001,Yunoki_1998}.

Recently, increasing interest has been devoted to 
the effect of geometrical frustration of the lattice structure 
on the charge-spin coupled systems. 
This has been inspired by the discovery of peculiar transport properties, 
such as an unconventional anomalous Hall effect, in some pyrochlore oxides~\cite{Taguchi2001,Nakatsuji2006,Machida2007}. 
Theoretically, it was shown that, 
in the ferromagnetic Kondo lattice model on a two-dimensional kagome lattice, 
the coupling to a noncoplanar spin ordering 
leads to the anomalous Hall effect via the so-called Berry phase 
from a finite spin scalar chirality~\cite{Ohgushi}. 
The idea was extended to other frustrated lattices, 
such as a face-centered-cubic lattice~\cite{Shindou2001}, 
a triangular lattice~\cite{M.B.,Akagi_2010}, and 
a pyrochlore lattice~\cite{Ikoma2003,Chern_2010}.
Among them, in the triangular-lattice systems that we focus on in the present study, 
it was pointed out that 
a perfect nesting of the Fermi surface at 3/4 electron
filling might lead to a noncoplanar four-sublattice ordering and the anomalous Hall effect~\cite{M.B.}. 
The authors performed careful energy comparison  
among possible spin orderings, and 
found that a noncoplanar four-sublattice spin ordering with a finite spin scalar chirality emerges 
near 1/4 filling in a wider parameter range than in 
the 3/4 filling case~\cite{Akagi_2010}. 
The 1/4 filling phase was later confirmed by Monte Carlo simulation~\cite{Kumar_Brink_2010,Kato_2010}. 
Such triangular-lattice charge-spin coupled systems have also drawn attention experimentally, 
e.g., in some delafossite oxides~\cite{Okuda_2005,Okuda_2008,Takatsu2009,Takatsu_2010}.

In this contribution, we extend the analysis of 
the ferromagnetic Kondo lattice model on the triangular lattice  
with focusing on the lightly-doped region. 
As mentioned above, the geometrical frustration brings about new aspects 
in the charge-spin coupled systems, 
but the keen competition between the AF SE and the ferromagnetic DE interactions 
in the lightly-doped region has not been studied in detail. 
On the triangular lattice, the AF SE interaction stabilizes 
a three-sublattice 120$^\circ$ AF order when the electron band is empty or fulfilled. 
We clarify the effect of both electron and hole doping to the 120$^\circ$ AF states 
by a variational calculation of the ground-state energy for various spin states  
up to four-sublattice ordering. 
In the electron doped case, we find that 
the 120$^\circ$ order is immediately collapsed by doping and 
a three-sublattice spin canting phase appears with accompanying PS. 
In contrast to the unfrustrated cases, this canting state has 
a noncoplanar umbrella-type spin configuration, which shows 
a finite scalar spin chirality together with a net magnetic moment.
On the other hand, in the hole doped case, 
the 120$^\circ$ AF state persists up to a finite doping rate and 
PS takes place to a ferromagnetic metal; 
a noncoplanar canting phase is not observed. 
The contrastive behavior between the electron and hole doping is discussed 
from the change of the density of states 
for the spin ordering patterns.

\section{Model and method}

We consider the ferromagnetic Kondo lattice model on the triangular lattice.
The Hamiltonian is given by 
\begin{equation}
{\cal H}=-t\sum_{\langle i,j \rangle,\alpha } ( c^{\dagger}_{i,\alpha } c_{j,\alpha }+\mathrm{h.c.}) 
-J_{\rm H} \sum_{i,\alpha ,\beta }
    c^{\dagger}_{i,\alpha } \vec {\sigma}_{\alpha \beta }  c_{i,\beta } \cdot \vec {S_i} 
   +J_{\rm K} \sum_{\langle i,j\rangle} \vec {S_i} \cdot \vec {S_j},
   \label{eq:H}
\end{equation}
where $c^{\dagger}_{i,\alpha }$ ($c_{i,\alpha }$) is a creation (annihilation) operator for a conduction electron with spin $\alpha$ at site \textit{i}, 
$\vec {\sigma}_{\alpha \beta }=({\sigma}^x_{\alpha \beta },{\sigma}^y_{\alpha \beta },{\sigma}^z_{\alpha \beta })$ is a vector of Pauli matrices, and 
$\vec {S_i}$ is a localized spin on site \textit{i}. 
Here, \textit{t} is the transfer integral between the nearest-neighbor sites on the triangular lattice, $J_{\rm H}$ is the Hund's-rule coupling between conduction electrons and localized spins, and $J_{\rm K}$ is the antiferromagnetic (AF) superexchange (SE) interaction between localized spins.  
We take $t=1$ as an energy unit and consider classical spins for $\vec {S_i}$ with $|\vec {S_i}|=1$.

We investigate the ground state of the model given by Eq.~(1) 
following the variational method taken in Ref.~\cite{Akagi_2010}. 
Namely, we compare the ground-state energies for different ordered states of the localized spins and determine the most stable ordering, 
while varying the electron density $n=\frac{1}{N}\sum_{i\alpha }\langle c^{\dagger}_{i,\alpha } c_{i,\alpha }\rangle$ (\textit{N} is the total number of sites), $J_{\rm H}$, and $J_{\rm K}$. 
In the calculation, we consider 13 different types of ordered states, up to a four-sublattice unit cell, as shown in Fig.~1 in Ref.~\cite{Akagi_2010}. 
For the states with spin canting, we optimize the canting angle $\theta $.

\section{Result and Discussion}

\begin{figure}[h]
\includegraphics[width=22pc]{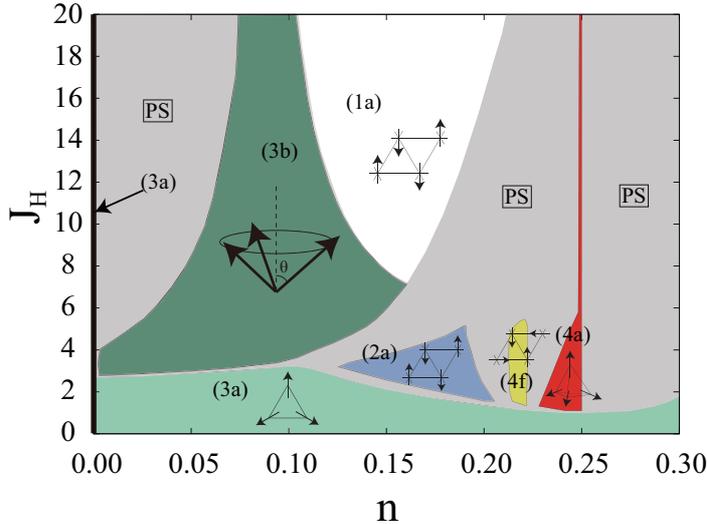}\hspace{2pc}
\begin{minipage}[b]{14pc}\caption{\label{phase-diagram}Ground-state phase diagram as functions of \textit{n} and $J_{\rm H}$ at $J_{\rm K}=0.03$.
Labels for the phases are common to those in Ref.~\cite{Akagi_2010}.
(3b) is the umbrella-type noncoplanar phase appearing in between 
the ferromagnetic metallic state (1a) and the phase separation (PS).
The thick black line at $n=0$ indicates that 
(3a) $120^{\circ }$ coplanar ordered phase is stabilized at zero doping.
}
\end{minipage}
\end{figure}

\begin{figure}[h]
\includegraphics[width=38pc]{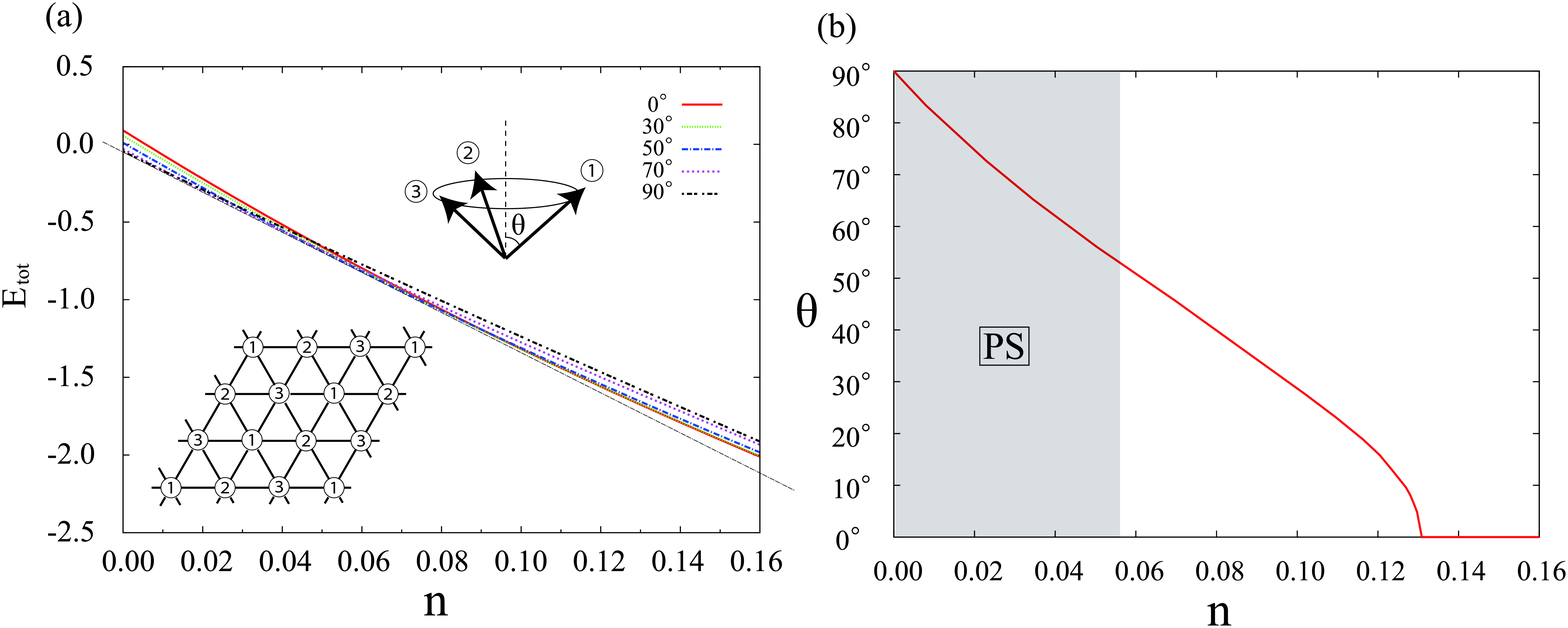}\hspace{2pc}
\begin{minipage}[b]{38pc}\caption{\label{total-energy}
(a) 
Energy per site of the umbrella-type noncoplanar ordered state. 
Different curves represent the energies 
for a fixed different canting angle $\theta$. 
The data are calculated at $J_{\rm K}=0.03$ and $J_{\rm H}=10$. 
Note that 
$\theta =0^{\circ }$ and $\theta =90^{\circ }$ are equivalent to (1a) ferromagnetic phase and (3a) three-sublattice 120$^\circ$-ordered phase, respectively.
The straight black line 
shows a tangent line 
to define the phase-separated region. 
The insets show schematic pictures of the spin configuration of the umbrella state 
and the triangular lattice. 
(b) $n$ dependence of $\theta$ which gives the lowest energy. 
The gray region denotes the phase-separated region. 
See the text for details. 
}
\end{minipage}
\end{figure}

Figure~\ref{phase-diagram} shows the result of the ground-state phase diagram at $J_{\rm K}=0.03$ 
in the lightly-doped region near $n=0$.
For comparison, the results at smaller $J_{\rm K}$ are found
in Fig.~3 in Ref.~\cite{Akagi_2010}. 
A new feature emerges in the low-density region in Fig.~\ref{phase-diagram}, 
which was not seen in the previous results for smaller $J_{\rm K}$.
That is, a spin-canting phase (3b) appears on the lower-$n$ side of 
the ferromagnetic metallic phase (1a). 
This state has a three-sublattice ordering with an umbrella-type noncoplanar spin configuration, 
as schematically shown in the figure [see also the inset of Fig.~\ref{total-energy}(a)]. 
Spin canting itself is expected in general in lightly-doped region
because of the competition between the AF SE interaction $J_{\rm K}$ 
and the ferromagnetic DE interaction~\cite{de-Gennes_1960}. 
For example, in the cubic lattice case, a plausible canting state, 
which naturally connects two-sublattice collinear AF state and doped ferromagnetic state, 
is a simple coplanar spin-flop state.
However, it is not trivial 
what type of spin-canting 
order is realized in geometrically-frustrated systems. 
In the present triangular-lattice case, there are several possibilities of different spin canting orders, 
as listed in Fig.~1 in Ref.~\cite{Akagi_2010}; 
even for the three-sublattice orders, there are three different types of canting states, 
(3b) umbrella, (3c) coplanar cant, and (3d) 2:1 cant (see Fig.~1 in Ref.~\cite{Akagi_2010}).
Among the possibilities, the umbrella-type one is energetically favored 
in the present model. 
Note that the umbrella state has a finite scalar spin chirality in each triangle unit, 
but this spatial configration does not lead to the anomalous Hall effect.

The umbrella phase is developed by increasing $J_{\rm K}$ in the region near 
the border of the ferromagnetic phase to PS as $J_{\rm K}$ increases. 
PS occurs to the $n=0$ state, in which itinerant electrons are absent 
and the localized spins are 120$^\circ$ ordered by $J_{\rm K}$. 
It is noteworthy that the umbrella phase is always accompanied by PS. 
The situation is demonstrated 
in Fig.~\ref{total-energy}. 
Figure~\ref{total-energy}(a) shows 
the energy per site at a fixed canting angle $\theta$ for various values of $\theta$. 
Here, 
$\theta =0^{\circ }$ and $\theta =90^{\circ }$ 
correspond to the 
ferromagnetic phase (1a) 
and 
the three-sublattice 120$^\circ$-ordered phase (3a), respectively. 
As $\theta$ decreases from 90$^\circ$, 
the energy at $n=0$ increases because of the loss of the SE energy. 
At the same time, the initial slope $dE/dn|_{n=0}$ becomes steeper 
because the kinetic energy 
of electrons becomes larger for smaller $\theta$ 
via the DE mechanism. 
Hence, the envelope of the energy curves $E(n,\theta)$ 
becomes convex upward in the low-density limit. 
The convex behavior is a sign of PS. 
The phase-separated region is determined by drawing a tangent line 
from the energy at $n=0$, $E(n=0,\theta=90^\circ)$, 
to the envelope of the energy curves $E(n>0,\theta<90^\circ)$. 
In the present case at $J_{\rm K}=0.03$ and $J_{\rm H}=10$, 
the tangent line touches the envelope at $n \simeq 0.06$ 
with the energy curve for $\theta \simeq 50^\circ$; 
i.e., the system exhibits PS 
between the 120$^\circ$-ordered state at $(n,\theta)=(0,90^\circ)$ and 
the umbrella-type canting state at $(n,\theta) \simeq (0.06,50^\circ)$. 
Similar behavior is observed for all the regions of the umbrella phase, 
which accounts for the reason 
why the umbrella phase always appears with accompanied by PS.

The canting angle $\theta$ which gives the lowest-energy envelope 
changes continuously in the umbrella phase, 
as seen in Fig.~\ref{total-energy}(a). 
Figure~\ref{total-energy}(b) gives a plot of the optimized $\theta$ as a function of $n$. 
The result shows that, as $n$ increases, 
$\theta$ starts from about $50^\circ$ at the border of PS 
at $n\simeq 0.06$, 
and gradually decreases to zero 
when approaching the boundary to the ferromagnetic phase at $n \simeq 0.13$. 

\begin{figure}[h]
\includegraphics[width=20pc]{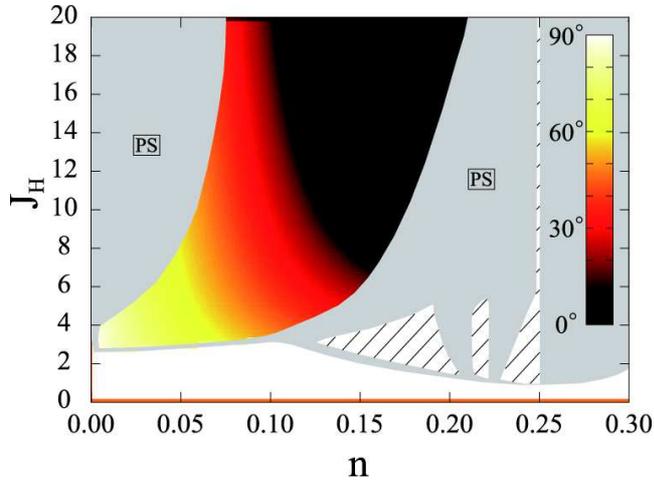}\hspace{2pc}
\begin{minipage}[b]{16pc}\caption{\label{angular-phase-diagram}
Canting angle $\theta$ in the umbrella phase 
as functions of \textit{n} and $J_{\rm H}$ at $J_{\rm K} =0.03$.
The ferromagnetic phase ($\theta=0^{\circ}$) and the 120$^\circ$-ordered phase ($\theta=90^{\circ}$) 
are also shown by black and white, respectively, whereas other phases are omitted 
(the hatched areas). 
The gray areas are the phase separation. 
}
\end{minipage}
\end{figure}

The value of the optimized $\theta$ depends not only on $n$ but also on $J_{\rm H}$ (and $J_{\rm K}$). 
The behavior is summarized in 
Fig.~\ref{angular-phase-diagram} at $J_{\rm K}=0.03$.
As shown in the figure, $\theta$ decreases as $n$ and $J_{\rm H}$ increase. 
This is reasonable since larger $n$ and $J_{\rm H}$ enhance 
the DE ferromagnetic interaction which favors smaller canting angle $\theta$ 
(larger net magnetic moment). 
This change of $\theta$ clearly demonstrates that the noncoplanar umbrella phase is stabilized 
in the competition between the AF SE interaction $J_{\rm K}$, 
which is dominant at less carrier doping and small $J_{\rm H}$, and 
the DE ferromagnetic interaction enhanced by doping and $J_{\rm H}$. 

In contrast to the above results for the electron doped case, 
a hole doping from the band top $n=1$ 
does not lead to a noncolpanar canting state. 
In the hole-doped case, 
120$^\circ$ coplanar ordering 
persists up to a finite doping concentration, and 
PS takes place between the doped 120$^\circ$ AF metal and a ferromagnetic metal
when $J_{\rm H}$ is sufficiently large. 
The situation is unchanged against the value of $J_{\rm K}$; 
a typical phase diagram for $n \sim 1$ is found 
in the result at $J_{\rm K} = 0.01$ in Fig.~3 in Ref.~\cite{Akagi_2010}.

\begin{figure}[h]
\includegraphics[width=20pc]{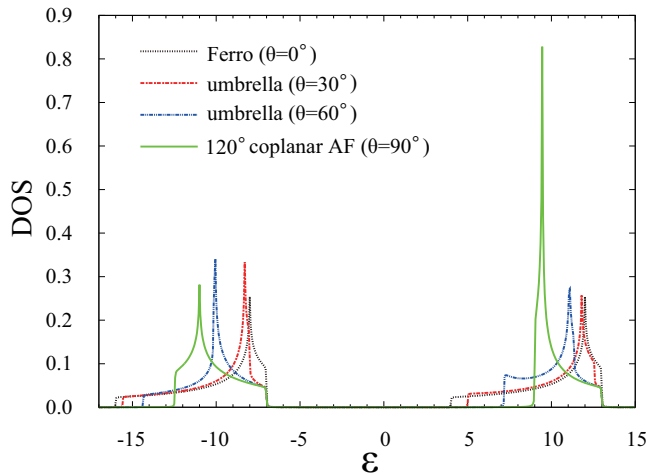}\hspace{2pc}
\begin{minipage}[b]{16pc}\caption{\label{DOS_JH8}
The density of states (DOS) for the umbrella-type canted states 
including the 120$^{\circ }$ coplanar AF state and 
the ferromagnetic state. 
The data are calculated at $J_{\rm H}=10$ and $J_{\rm K}=0$.
}
\end{minipage}
\end{figure}

The reason why the 120$^\circ$ AF order is stable against the hole doping is understood from the density of states (DOS). 
Figure~\ref{DOS_JH8} shows DOS for the 120$^{\circ }$ coplanar AF phase,  
in comparison with typical DOS for the umbrella canting states including
the ferromagnetic state. 
The data are calculated at $J_{\rm H}=10$ and $J_{\rm K}=0$.
The result shows that the band top energy $\varepsilon = 13t$ is unchanged for $\theta$.   
This is confirmed by analytical calculations. 
In particular, a consideration in the limit of $J_{\rm H} \to \infty$ provides 
simple understanding of the form of DOS 
in terms of the renormalization of the transfer integrals and 
the flux introduced by the Berry phase~\cite{Anderson1955,Wang2005}. 
While the band top energy is unchanged, the value of DOS at the band top 
is smaller for $\theta > 0$ compared to the $\theta = 0$ ferromagnetic state. 
For these situations, the kinetic energy by hole doping becomes lower for $\theta > 0$ states.  
Among the $\theta > 0$ states, the 120$^\circ$ AF state has the lowest AF SE energy, and 
therefore, the 120$^\circ$ AF ordering remains stable against the hole doping 
for $J_{\rm K} > 0$.

DOS in Fig.~\ref{DOS_JH8} also accounts for the reason 
why electron doping from $n=0$ immediately collapses the 120$^\circ$ AF state and 
causes PS to a canted state. 
As shown in the figure, the band-bottom energy is lower for smaller $\theta$, and hence, 
the electron doping to the 120$^\circ$ AF state costs the energy than the canted states. 
At $n=0$, the 120$^\circ$ AF ordering becomes stable 
by a small energy gain from the AF SE coupling, $-3 J_{\rm K}$ per site. 
The energy gain is compensated by infinitesimally small doping 
to a canted state with smaller $\theta$ because of the lower band bottom. 
Therefore the electron doping leads to a sudden PS to a canted state, as demonstrated above.

The contrastive behavior between electron and hole doping is expected to 
be observed around an insulating state at half filling $n=0.5$. 
At $n=0.5$, the 120$^\circ$ coplanar AF ordering is stable and 
the system becomes insulating for $J_{\rm H}\hspace{0.3em}\raisebox{0.4ex}{$>$}\hspace{-0.75em}\raisebox{-.7ex}{$\sim $}\hspace{0.3em}2$ (see Fig.~3 in Ref.~\cite{Akagi_2010}). 
In the insulating state for larger $J_{\rm H}$, 
DOS just below and above the gap behave similar to the band top and bottom, respectively, 
as shown in Fig.~\ref{DOS_JH8}. 
In fact, as a preliminary result, we find a similar contrastive behavior around $n=0.5$; 
a noncoplanar canting state appears against electron doping with PS, 
while the 120$^\circ$ AF order retains up to a finite hole doping. 
The details will be reported elsewhere.

Finally, let us discuss briefly the effect of the band structure on these properties 
by including the next-nearest-neighbor hopping $t'$ in the model (\ref{eq:H}). 
We find that, when $t'$ has an opposite sign to $t$, 
$\theta$ dependence of DOS at the band edges is qualitatively unchanged. 
For $t'$ with the same sign as $t$, the band top becomes higher for smaller $\theta$, 
while the band bottom remains lower for smaller $\theta$, at least, for small $t'$. 
Hence, we expect the similar results 
for the models including $t'$ irrespective of its sign. 
Further analyses including the detailed energy comparison among different spin states 
will be reported elsewhere.

\section{Summary}

We have investigated the lightly-doped regions of 
the ferromagnetic Kondo lattice model on the triangular lattice by variational calculations. 
We have found that a three-sublattice noncoplanar phase is stabilized 
by electron doping from the band bottom. 
The noncoplanar spin canting phase has an umbrella-type spin configuration 
with a finite spin scalar chirality in each triangular plaquette. 
This phase always appears accompanied by a phase separation to 
the 120$^\circ$ coplanar antiferromagnetic insulating state at zero doping. 
On the contrary, hole doping from the band top does not lead to 
such a noncoplanar spin canting state. 
These peculiar properties are the consequence of geometrical frustration 
in the charge-spin coupled system in which the antiferromagnetic superexchange interaction and 
the ferromagnetic double-exchange interaction compete with each other.

We acknowledge helpful discussions with Takahiro Misawa, Masafumi Udagawa, Youhei Yamaji, and Junki Yoshitake. 
This work was supported by Grants-in-Aid for Scientific Research 
(No. 19052008 and 21340090), 
Global COE Program ``the Physical Sciences Frontier", the Next Generation Super Computing 
Project, and Nanoscience Program, from MEXT, Japan.

\end{document}